\documentstyle[13lomcon,cite,epsfig]{article}

\begin{document}

\title{BETA-BEAMS}

\author{ C. Volpe \footnote{e-mail: volpe@ipno.in2p3.fr}}

\address{Institut de Physique Nucl\'eaire Orsay, F-91406 Orsay cedex, FRANCE}

\maketitle\abstracts{ Beta-beams is a new concept for the production of intense and
pure neutrino beams. It is at the basis of a proposed
neutrino facility, whose main goal is to explore the possible existence of CP
violation in the lepton sector. Here we briefly review the original
scenario and the 
low energy beta-beam. This option would offer a unique opportunity to perform neutrino
interaction studies of interest for particle physics,
astrophysics and nuclear physics. Other proposed
scenarios for the search of CP violation are mentioned.}

\section{Introduction}

The observations made by the Super-Kamiokande \cite{fukuda},
the K2K \cite{ahn}, the SNO \cite{ahmad} and
the KAMLAND \cite{eguchi} 
experiments have brought a breakthrough in the field of neutrino physics.
The longstanding puzzles of the solar neutrino deficit \cite{davis} and
of the atmospheric anomaly
have been clarified :
the expected fluxes are reduced due to the neutrino oscillation phenomenon, i.e.
the change in flavour that neutrinos undergo while traveling \cite{pontecorvo}. 
The overall picture is 
now also confirmed by the recent mini-BOONE result \cite{AguilarArevalo:2007it}. 

Neutrino oscillations imply that neutrinos are
massive particles and represent
the first direct experimental evidence for physics beyond the Standard Model. 
Understanding the mechanism for
generating the neutrino masses and their small values is clearly a fundamental
question, that needs to be understood. On the other hand, the presently known
(as well as unknown) neutrino properties have important implications for other
domains of physics as well, among which astrophysics, e.g. for our
comprehension of processes like the nucleosynthesis of heavy elements,
and cosmology. 

An impressive progress has been achieved in our knowledge of neutrino
properties. Most of the parameters
of the Maki-Nakagawa-Sakata-Pontecorvo (MNSP) unitary matrix \cite{mns}, relating the
neutrino flavor to
the mass basis, are nowadays determined, except
the third neutrino mixing angle, usually called $\theta_{13}$.
However, this matrix might be complex, meaning
there might be (one or more) phases. 
A non-zero Dirac phase
introduces a difference between neutrino and anti-neutrino oscillations and
implies the breaking of the $\cal{CP}$ symmetry in the lepton sector.
Knowing its value might require the
availability of very intense neutrino beams in next-generation accelerator neutrino
experiments, namely super-beams, neutrino
factories or beta-beams. Besides representing a crucial discovery, 
the observation of a non-zero phase 
might help unraveling the asymmetry
between matter and anti-matter in the Universe and have an impact in
astrophysics, e.g. for core-collapse supernova physics \cite{Balantekin:2007es}.

Zucchelli has first proposed the idea of producing electron (anti)neutrino
beams using the beta-decay of boosted radioactive ions: the ``beta-beam''
\cite{zucchelli}.
It has three main advantages: well-known fluxes, purity
(in flavour) and collimation. This simple idea exploits major developments in the 
field of nuclear physics, where radioactive ion beam facilities
under study such as 
the european EURISOL project 
are expected to reach ion intensities of $10^{11-13}$ per second.
A feasibility study of the original scenario is ongoing (2005-2009) 
within the EURISOL Design Study (DS) financed by the
European Community.

At present, various beta-beam scenarios can be found in the literature, depending
on the ion acceleration. They are usually classified following the value of the Lorentz
$\gamma$ boost factor, as low energy ($\gamma=6-15$)
\cite{Volpe:2003fi,McLaughlin:2003yg,Serreau:2004kx,McLaughlin:2004va,Balantekin:2005md,Balantekin:2006ga,Jachowicz:2006xx,Barranco:2007ej,Lazauskas:2007va,Lazauskas:2007bs,Amanik:2007ce,Amanik:2007ce,Amanik:2007zy,Bueno:2006yg,nathalie},
original ($\gamma \approx 60-100$) 
\cite{zucchelli,maur03,maur05,bur05,gugl05,cam06,ber05}, 
medium ($\gamma$ of several hundreds)
and high-energy ($\gamma$ of the order of thousands)
\cite{bur04,don05,hub06,Agarwalla:2006vf,Agarwalla:2006gz}. 
(For a review of all scenarios see \cite{Volpe:2006in}.)
An extensive investigation of the corresponding physics potential
is being performed and new ideas keep being proposed. For example, a radioactive ion beam
production method is discussed in \cite{rubbia06} and will be investigate within the new "EuroNU" DS.
Thanks to this method
two new ions $^{8}$B and $^{8}$Li are being considered as candidate emitters, 
while the previous literature is mainly focussed on
$^{6}$He and $^{18}$Ne. The corresponding sensitivity is
currently under study (see e.g. \cite{Coloma:2007nn}). 

\section{The original scenario}
\noindent
In the original scenario \cite{zucchelli}, the ions are produced,
collected, accelerated up to
several tens GeV/nucleon - after injection in the Proton Synchrotron 
and Super Proton Synchrotron accelerators at CERN -
and stored in a storage ring of 7.5 km (2.5 km) total length
(straight sections).
The neutrino beam produced by the decaying ions point to a large water 
\v{C}erenkov detector \cite{deBellefon:2006vq} (about 20 times Super-Kamiokande),
located at the (upgraded) Fr\'ejus Underground Laboratory, in order
to study $\cal{CP}$ violation, through a comparison of 
$\nu_e \rightarrow \nu_{\mu}$ and
$\bar{\nu}_e \rightarrow \bar{\nu}_{\mu}$ oscillations.
This facility is based on reasonable extrapolation 
of existing technologies
and exploits already existing accelerator infrastructure to reduce cost.
Other technologies are being considered 
for the detector as well \cite{Autiero:2007zj}.
A first feasibility study is performed in \cite{autin}. 

The choice of the candidate emitter has to meet several criteria, including a high intensity
achievable at production and a not too short/long half-life. 
The ion acceleration window is determined by a compromise between having the
$\gamma$ factor as high as possible, to profit of larger cross sections
and better focusing of the beam on one hand, and keeping it as low as
possible to minimize the single pion background and better match the
$\cal{CP}$ odd terms on the other hand. 
The signal corresponds to the muons produced by
$\nu_{\mu}$ charged-current events in water, mainly via quasi-elastic
interactions at these energies. Such events are selected by requiring a
single-ring event, with the same identification algorithms used by the
Super-Kamiokande experiment, and by the detection
of the electron from the muon decay.
At such energies the energy resolution is very poor due
to the Fermi motion and other nuclear effects. For these reasons, a $\cal{CP}$
violation search with $\gamma=60-100$ is based on a counting experiment only.
\begin{table}
\caption{\label{t:events}
Number of events expected after 10 years, 
for a beta-beam produced at CERN and sent to 
a 440 kton water \v{C}erenkov detector located at an (upgraded) Fr\'ejus Underground Laboratory, at 130 km 
distance. The results correspond to $\bar{\nu}_e$ (left)  and  $\nu_e$ (right). 
The different $\gamma$
values are chosen to make the ions circulate together in the ring \cite{maur05}}
\begin{tabular}{@{}lrr}
&  $^6$He               &   $^{18}$Ne                \\
&       ($\gamma=60$)  &  ($\gamma=100$)  \\
\hline
CC events (no oscillation) & 19710   & 144784  \\
Oscillated ($\sin^2 2\theta_{13}=0.12$, $\delta=0$) & 612 & 5130   \\
Oscillated ($\delta=90^\circ$,$\theta_{13}=3^\circ$)&  44     &  529   \\
Beam background            &  0      &  0\\
Detector backgrounds       &   1     &  397\\
\end{tabular}
\end{table}

The beta-beam has no intrinsic
backgrounds, contrary to conventional sources. However, inefficiencies in
particle identification, such as single-pion production in neutral-current
$\nu_e$ ($\bar{\nu}_e$) interactions, electrons (positrons) misidentified
as muons, as well as external sources, like atmospheric neutrino interactions,
can produce backgrounds. The background coming from single pion production
has a threshold at about 450 MeV, therefore giving no contribution for
$\gamma < 55$. Standard algorithms for particle identification
in water \v{C}erenkov detectors are quite efficient in suppressing the fake
signal coming from electrons (positrons) misidentified as muons.
Concerning the atmospheric neutrino interactions, 
estimated to be of about 50/kton/yr, this important background is reduced to
1 event/440 kton/yr by requiring a time bunch length for the ions of 10 ns.
The expected events from \cite{maur05} are shown in Table \ref{t:events}, as an example.

\begin{figure}[t]
  \begin{minipage}{5cm}
   \centering
     \includegraphics[scale=0.45]{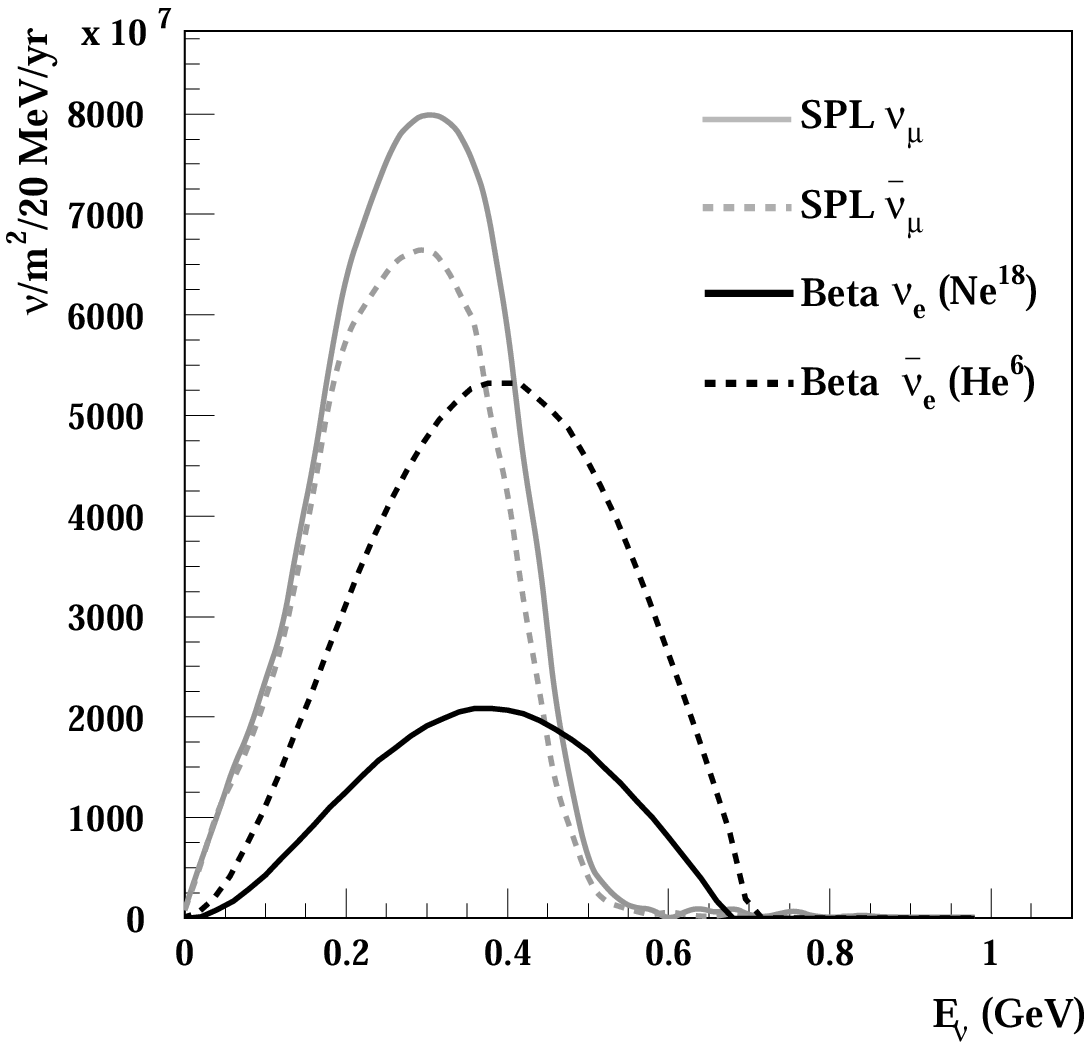}
     \caption{Comparison of neutrino fluxes from a super-beam (SPL) and a beta-beam. 
The ions circulate at the same $\gamma=100$, independently, in the storage ring.
Note that the average neutrino energies are related to the ion boost through
$E_{\nu} \approx 2 \gamma Q_{\beta}$ \cite{gugl05}. }
     \label{fig:fluxst}
  \end{minipage}
\hfill
  \begin{minipage}{5cm}
     \includegraphics[scale=0.31]{CP-systematics.eps}
     \caption{$\cal{CP}$ 
   discovery reach for
  the ($\gamma=100$) beta-beam ($\beta$B), a super-beam (SPL), 
  and T2HK as a function of $\theta_{13}$. The width of the bands corresponds
  to values with 2\% to 5\% systematical errors. 
 \cite{cam06}. }
     \label{fig:CP}
  \end{minipage}
\end{figure}
The discovery potential is analyzed in
\cite{zucchelli,maur03,maur05,bur05,gugl05,cam06,ber05}.
A detailed study of $\gamma=100$ option is made 
for example in \cite{cam06}
based on the GLoBES software \cite{hub05},
including correlations and degeneracies and using atmospheric data in the analysis
\cite{hub06}. The fluxes are shown in Figure \ref{fig:fluxst}. 
Figure \ref{fig:CP} shows the $\cal{CP}$  
discovery reach as an example of the sensitivity that can be reached 
running the ions around $\gamma=100$.

\section{Low energy beta-beams}
\begin{figure}[tbp]
\begin{center}
\includegraphics[scale=0.5]{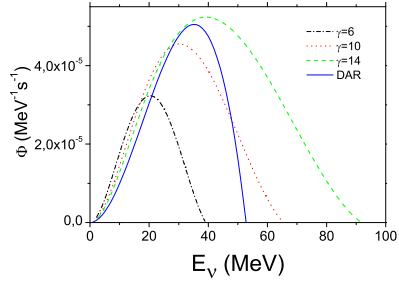}
\end{center}
\caption{Anti-neutrino fluxes from the decay of $^{6}$He ions boosted at $%
\protect\gamma=6$ (dot-dashed line),$\protect\gamma=10$ (dotted line) and $%
\protect\gamma=14$ (dashed line). The full line presents the Michel spectrum for
neutrinos from muon decay-at-rest. }
\label{fig:lownu}
\end{figure}

\begin{figure}[t]
\centering
\includegraphics[scale=0.2]{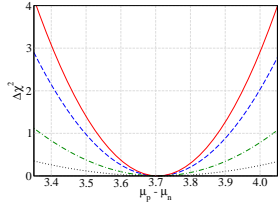}
\caption{CVC Test : $\Delta \chi^2$ obtained from the angular distribution of electron anti-neutrinos on proton scattering in a water \v{C}erenkov detector in the cases when the statistical error only (solid), with 2\% (dashed), 5\% (dash-dotted) and 10\% (dotted) systematic errors.  The 1$\sigma$ ($\Delta \chi^2=1$) relative uncertainty in the weak magnetism contribution $\mu_p - \mu_n$ is 4.7\%, 5.6\%, 9.0\% and more than 20\%, respectively.  The results correspond to $\gamma=12$ \cite{Balantekin:2005md}. }
\label{fig:CVC}
\end{figure}

\begin{figure}[t]
\centering
 \includegraphics[scale=0.2]{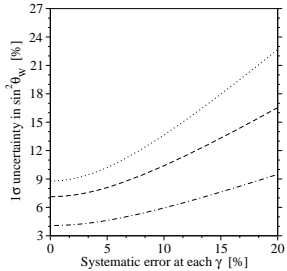}
\vspace{.1cm}
     \caption{Weinberg angle : One signa uncertainty as a function of the systematic error at each $\gamma$
for $\gamma=12$ (dotted),  $\gamma=7,12$ (broken) and $\gamma=7,8,9,10,11,12$ (dashed-dotted line). The results are obtained considering electron anti-neutrino scattering on electrons in a water \v{C}erenkov detector\cite{Balantekin:2006ga}.}
     \label{fig:wein}
\end{figure}

A low energy beta-beam facility producing neutrino beams 
in the 100 MeV energy range has been first proposed in \cite{Volpe:2003fi}.
Figure \ref{fig:lownu} shows the corresponding fluxes. 
The broad physics potential of such a facility, currently being analyzed,
covers :
\begin{itemize}
\item neutrino-nucleus interaction and nuclear structure studies 
\cite{Volpe:2003fi,Serreau:2004kx,McLaughlin:2004va,Lazauskas:2007bs};
\item electroweak tests of the Standard Model, such as 
a new method to test the Conserved-Vector-Current hypothesis \cite{Balantekin:2005md} (Figure \ref{fig:CVC}), a measurement
of the Weinberg angle at small momentum transfer \cite{Balantekin:2006ga} (Figure \ref{fig:wein}) or of 
the neutrino magnetic moment \cite{McLaughlin:2003yg});
\item core-collapse supernova physics \cite{Volpe:2003fi,Jachowicz:2006xx}.
\end{itemize}

Here I briefly mention some of the results concerning
the physics potential of a low energy beta-beam. 
Neutrino-nucleus interaction is a topic of 
current great interest since the corresponding 
cross sections are necessary for the interpretation
of neutrino detector response, for the understanding of the nucleosynthesis of
heavy elements and in the search for physics beyond the Standard Model. 
For example, a precise knowledge of the neutrino-lead cross section
\cite{Volpe:2001gy} 
can be exploited to extract information on the third neutrino miwing angle
that still remains unknown \cite{Engel:2002hg}.
In spite of the numerous applications, the experimental
information is scarce since only a few experiments have been performed. The best studied
case, i.e. interactions on carbon still suffer from important discrepancies
between experiment and theory (see e.g.\cite{Volpe:2000zn}). 
The theoretical calculations of the cross sections for neutrino energies in the several
tens MeV are subject to nuclear structure uncertainties due to the different choices of
the approaches and effective interactions used.
(For a review see for example \cite{Volpe:2004dg,Balantekin:2003ip,Kolbe:2003ys}.)

In \cite{Serreau:2004kx} it is first shown that a devoted smaller storage ring might indeed be necessary to perform
neutrino-nucleus measurement at a low energy beta-beams facility,
since the experiments require a close detector and only the ions close to the ends of the
straight sections contribute. 
Alternatively, the low energy neutrino fluxes might be obtained by putting one/two detectors at off-axis of
the storage ring planned for the CP violation search \cite{Lazauskas:2007va}.  
A comparison of the physics potential of low energy
beta-beams and conventional sources is made in \cite{McLaughlin:2004va}. In \cite{Lazauskas:2007bs} 
it is shown that
interesting information on the spin-dipole as well as higher multipoles might be obtained by 
the vaying the $\gamma$
of the ions. Instead of varying the ions one might take different parts of the fluxes at a detector 
\cite{Amanik:2007zy}.
Gathering more experimental constraints on the
corresponding transition amplitudes is important since the same
nuclear matrix elements are involved in the neutrinoless double-beta decay due to exchange of
a massive Majorana neutrino \cite{Volpe:2005iy}. These measurements can furnish a new constraints
to the half-lives calculations that are still plagued by important discrepancies.
Finally, a combination of the neutrino beams at different boost can be used to reconstruct
the signal from a supernova explosion in an observatory \cite{Jachowicz:2006xx}. The proposed method
has the advantage that it is free from the cross section uncertainties. 

\section{The other scenarios for CP violation searches}

Various scenarios for the study of $\cal{CP}$ violation
have been proposed where the energy of the
ions is much higher, the $\gamma$ ranging from
150 \cite{bur05} to several hundreds
to thousands \cite{bur04,don05,hub06,Agarwalla:2006vf,Agarwalla:2006gz,Coloma:2007nn}. 
The value of 150 GeV per nucleon comes from the maximum
acceleration that can be attained in the SPS. 
The baseline scenario in this case is the same as the original one.
On the contrary, the medium and high energy options
require major changes in the accelerator infrastructure, such as a 
refurbished SPS (or even the LHC) at CERN, as well as
bigger storage rings.  
To match the same oscillation frequencies, such scenarios need
further locations for the far detector, such as the Canfranc or the Gran
Sasso Underground Laboratories. 
The physics potential covers the third neutrino
mixing angle, the $\cal{CP}$ violating phase, as well as the neutrino mass
hierarchy. Some reduction of the degeneracy problem is also expected.
A specific feasibility study is still to be done in order
to determine e.g. the ion intensities (that drastically influence the
sensitivities) and the storage ring characteristics.

Monochromatic neutrino beams produced by boosted 
ions decaying through electron capture
are proposed in \cite{ber05}.
The baseline envisaged is the same as for the original beta-beam.
A comparison of $\nu_e \rightarrow \nu_{\mu}$ 
oscillations (only $\nu_e$ are available) at different neutrino energies is necessary.
Such a configuration requires the acceleration and storage 
of not fully stripped ions. The achievable ion rates need to be determined.

\section{Conclusions}
A beta-beam facility has
a rich and broad physics potential. The future and ongoing 
feasibility studies as well as the current physics investigation 
will furnish the necessary elements to
assess the final CP violation discovery reach. 
On the other hand, the availability of low energy beta-beams  
would open new research axis of interest for
particle, nuclear and core-collapse supernova physics. This option might require either
a devoted storage ring or detector(s) at off-axis.

\end{document}